\documentclass[aps,prb,twocolumn,superscriptaddress,floatfix,10pt]{revtex4-2}
\usepackage{caption}
\usepackage[english]{babel}
\usepackage{caption}
\usepackage{amsmath,amssymb} 
\usepackage{bm} 
\usepackage{graphicx} 
\usepackage{subcaption} 

\usepackage{comment} 

\usepackage{siunitx}

\usepackage{minted} 
\usepackage{textcomp} 
\usepackage{xcolor}

\usepackage{enumitem}
\setlist{noitemsep,leftmargin=*,topsep=0pt,parsep=0pt}

\usepackage{xcolor} 
\definecolor{lightgray}{gray}{0.6}
\definecolor{medgray}{gray}{0.4}

\usepackage{hyperref}
\hypersetup{
colorlinks=true,
urlcolor= blue,
citecolor=blue,
linkcolor= blue,
}

\newif\ifptitle
\newif\ifpnumber
\newcounter{para}

\ptitletrue  
\pnumbertrue  

\newcommand{\mytitle}{Detection of Image Potential States above the vacuum level in GeTe}

\usepackage{cleveref} 
\begin{document}

\title{\mytitle}

\author{Frédéric Chassot}
\email[]{frederic.chassot@unifr.ch}
\affiliation{Department of Physics and Fribourg Center for Nanomaterials, Université de Fribourg, Fribourg, Switzerland}

\author{Aki Pulkkinen}
\affiliation{New Technologies-Research Center,  University of West Bohemia in Pilsen, Plzeň, Czech Republic}

\author{Ján Minár}
\affiliation{New Technologies-Research Center,  University of West Bohemia in Pilsen, Plzeň, Czech Republic}

\author{Gunther Springholz}
\affiliation{Institut für Halbleiter-und Festkörperphysik, Johannes-Kepler-Universität, Linz, Austria}

\author{Matthias Hengsberger}
\affiliation{Physik-Institut, Universität Zürich, Zürich, Switzerland}

\author{Claude Monney}
\email[]{claude.monney@unifr.ch}
\affiliation{Department of Physics and Fribourg Center for Nanomaterials, Université de Fribourg, Fribourg, Switzerland}

\date{\today}

\begin{abstract}
The ferroelectric semiconductor $\alpha$-GeTe(111) has attracted significant attention in the last decade due to its unique properties, with extensive studies focusing on its occupied electronic bandstructure. In contrast, its unoccupied states - particularly those near the conduction band minimum - remain largely unexplored. In an effort to characterize those states, we surprisingly observe three image-potential states (IPS) in $\alpha$-GeTe(111) extending up to 0.8 eV above the vacuum level. Using time- and angle-resolved photoemission spectroscopy, we resolve the full parabolic dispersions of the first three IPS and determine their binding energies. Our analysis, combined with Bloch spectral function calculations, reveals that the unexpected persistence of IPS above the vacuum level originates from strong dipole transitions and the presence of large electron reservoirs in GeTe. 

\end{abstract}

\maketitle
\section{\label{sec:Intro}Introduction}

$\alpha$-GeTe(111) is a fascinating material that combines semiconducting properties with ferroelectricity, making it an attractive platform for both fundamental studies and technological applications. Below 670 K, a spontaneous lattice distortion induces a macroscopic polarization \cite{pawley_diatomic_1966}. The broken inversion symmetry leads to a momentum-dependent Rashba-type spin splitting of the bulk bands \cite{bychkov1984properties, Rotenberg_Spin-Orbit_1999}. Notably, GeTe exhibits one of the largest Rashba parameters reported to date, $\alpha_R\approx4.2$ eV \AA{} \cite{di_sante_electric_2013,krempasky_disentangling_2016}, making it interesting for spin-to-charge conversion \cite{rinaldi_evidence_2016, varotto_room-temperature_2021}, non-volatile information storage \cite{picozzi_ferroelectric_2014}, spin Hall effect generation \cite{wang_spin_2020}, and ultrafast control of ferroelectricity using femtosecond pulses \cite{kremer_field-induced_2022}. These prospects have driven intense research efforts over the past decade, particularly focused on its occupied bandstructure and spin texture \cite{krempasky_entanglement_2016, liebmann_giant_2016, elmers_spin_2016, rinaldi_ferroelectric_2018, krempasky_spin-resolved_2019, krempasky2020fully}.

Image-potential states (IPS) have been extensively studied since their prediction in the late 1970s \cite{echenique1978existence}. IPS are intermediate states in the 2-photon photoemission process after absorption of a first photon, corresponding to electrons bound outside of the material's surface by its dielectric response. These states exhibit nearly free-electron-like parabolic dispersions parallel to the surface with binding energies forming a discrete Rydberg series in the normal direction \cite{echenique1978existence}. Their photoemission requires absorption of a second photon. Initially predicted at metal surfaces, IPS were first confirmed experimentally by inverse photoemission \cite{straub1984intrinsic, dose1984image}. A deeper characterization of IPS became possible only with the advent of two-photon photoemission (2PPE) and energetically more precise photon sources, enabling detailed studies of their dispersion, binding energy, and spatial extension beyond noble metal surfaces \cite{hofer_time-resolved_1997,fauster_femtosecond_2000,echenique_decay_2004,wolf_ultrafast_1996,wolf_femtosecond_1997,link_femtosecond_2001,hertel_ultrafast_1996,rohleder_time-resolved_2005}. The tracking of  electron relaxation pathways provided further insights on the different interband and intraband scattering processes \cite{berthold_momentum-resolved_2002}.
While IPS have occasionally been reported in more recent works (see. e.g. Ref. \cite{sobota_ultrafast_2012}), they have received comparatively little attention in the context of semiconductors and ferroelectric materials. We emphasize here that IPS dispersions have been so far measured up to the vacuum level only, as they are not bound above.

In this work, we uncover a previously unexplored aspect of electronic structure of GeTe : the remarkable observation of IPS dispersion well above the vacuum level. Using time- and angle-resolved photoemission spectroscopy (TR-ARPES), we track the evolution of IPS across their full parabolic dispersions and characterize their behavior both below and above the vacuum level. Unlike 2PPE, ARPES provides direct access to the initial states, allowing us to identify the electron reservoirs that populate these intermediate states. Our findings reveal an unprecedented extension of IPS and provide insights into their formation mechanisms in a polar semiconductor.

\section{\label{sec:Methods}Methods}
\label{Section_Method}
Time- and Angle-resolved photoemission spectroscopy (TR-ARPES) were carried out using a Scienta DA30 photoelectron analyzer with a base pressure better than $3 \times 10^{-11}$ mbar. Static measurements were done using monochromatized He$_I$  radiation with $h\nu = 21.2$ eV. For time-resolved ARPES, half of the power of a femtosecond laser (Pharos, Light Conversion, operating at 1030 nm) is converted
into 780 nm (1.60 eV) light with an optical parametric amplifier, which is then frequency-quadrupled to 6.33 eV in $\beta-$BaB$_2$O$_4$ crystals
to generate UV pulses \cite{faure_full_2012}. The intrinsic resolution of the UV pulse is 25 meV,
as determined by the fit of the Fermi edge of a polycrystalline metal. The remaining half of the fundamental laser power is directed into a collinear optical parametric amplifier (Orpheus, Light Conversion) to generate IR pulses at 1.55, 1.38, 1.24 or 0.95 eV with a duration of about 80 fs. The temporal resolution was determined to be better than 100 fs by measuring the width of the photoemission cross-correlation between the pump and the probe pulses. The pump-probe measurements were performed at 200 kHz and with different IR fluences $F\in [127,226]$ $\mu$J/cm$^2$, with no significant difference between measurements at different fluences. We used p polarization for the IR and an incidence angle of 55°. The TR-ARPES data have been acquired with a negative bias voltage of -5 V applied to the sample. All the photoemission measurements were realized at 90~K if not further specified. The total energy resolution was about 45 meV and cooling of the sample was carried out at rates $<$5 K/min to avoid thermal stress. 

Sample Growth : Ferroelectric $\alpha$-GeTe films (500 nm thick) were grown by molecular beam epitaxy on (111) oriented BaF$_2$ substrates using a GeTe effusion cell and a growth temperature of 280 °C. Perfect 2D growth was observed by in situ reflection high-energy electron diffraction. After growth, the samples were transferred into a Ferrovac UHV suitcase in which they were transported for the TR-APRES measurements without breaking UHV conditions (pressure $<1\cdot 10^{-10}$ mbar).

Calculations:
The electronic structure of the Te-terminated GeTe(111) surface was calculated in the semi-infinite geometry using the screened Korringa-Kohn-Rostoker approach as implemented in the SPR-KKR package \cite{Ebert_2011}, and local spin density approximation (LSDA) within the Vosko-Wilk-Nusair parameterization of the exchange-correlation functional \cite{Vosko_1980}. The lattice parameters of GeTe were taken as $a=4.1735$ \AA, $c=10.692$ \AA, and $z_{\rm Te} = 0.4778$. Since the semilocal density functional theory is unable to describe the proper form of the potential barrier in the vacuum region, we have shifted the vacuum potentials to follow the $1/z$ form and set the image plane $0.52$ \AA{} outside of the outermost Te layer (see Ref. \cite{echenique2002image} for explanation of image plane). To account for the underestimation of band gap within LSDA, the unoccupied part of the band structure has been shifted by 0.38 eV.

\section{\label{sec:Results}Results}

\begin{figure}  \begin{center}\begin{subfigure}[b]{1\linewidth}
    \phantomcaption
    \label{fig:1a}
    \includegraphics[width=\columnwidth]{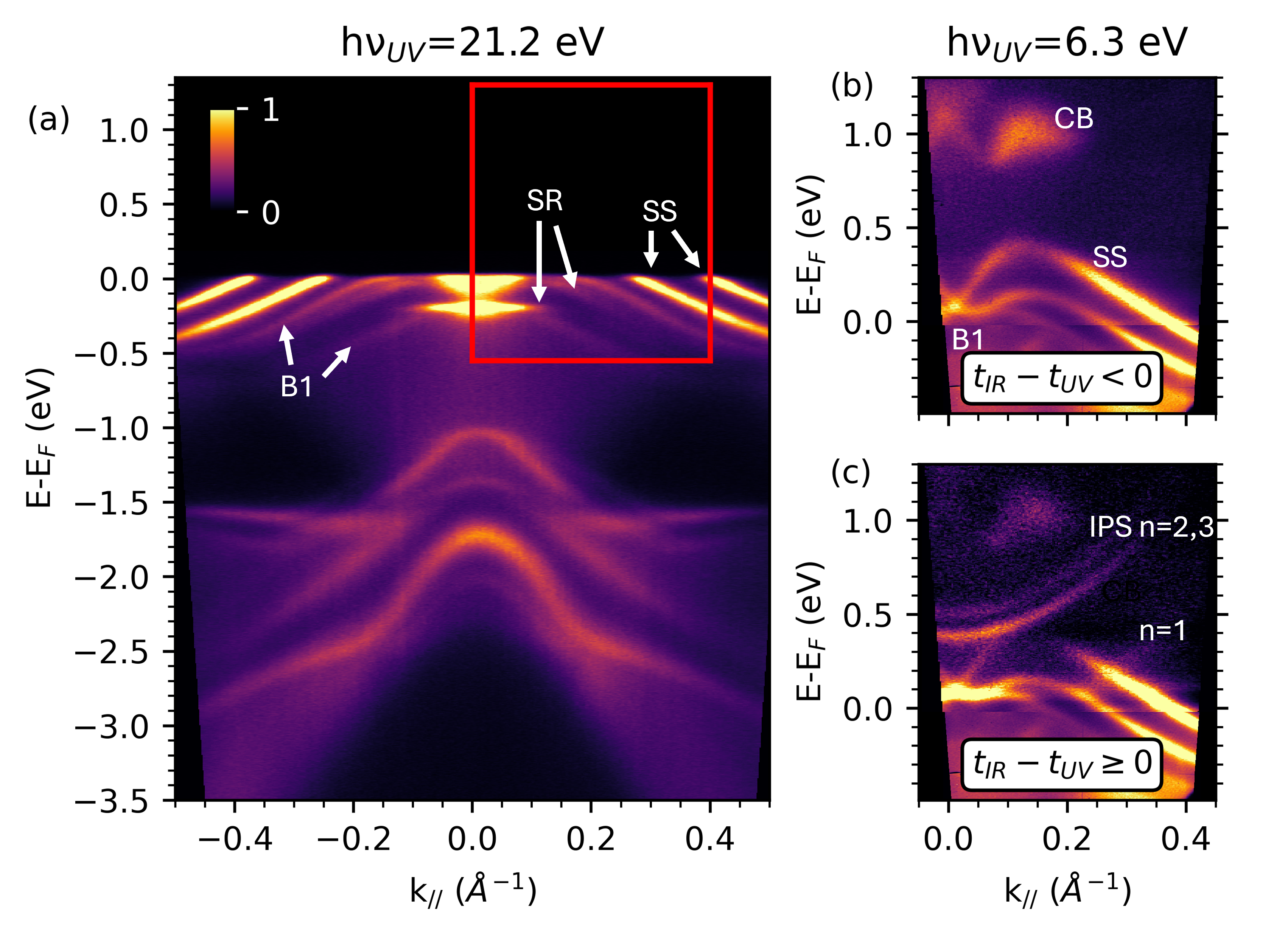}
    \phantomcaption
    \label{fig:1b}
    \phantomcaption
    \label{fig:1c}
   \end{subfigure}\end{center}
    \caption{\textbf{Bandstructure of $\alpha$-GeTe(111).} (a) ARPES measurements along $\overline{K\Gamma K}$ at 50 K with a probe photon energy of 21.2 eV with highlighted bulk (B1), surface resonance (SR) and surface states (SS). The photoemission intensity is plotted against the wavevector $k_{\parallel }$ and initial state energy $E-E_F$. (b) Snapshot of a TR-ARPES measurement along the $\overline{\Gamma K}$ direction (in the red area in (a)) at 90 K, taken with a UV photon energy of 6.3 eV and an IR photon energy of 1.55 eV obtained at negative IR-UV delays (integrated between $-50$ fs and $-250$ fs). (c) Same than (b) but integrated between $+50$ fs and $+275$~fs).}
    \label{fig:1}
\end{figure}

\begin{figure}  \begin{center}\begin{subfigure}[b]{1\linewidth}
    \includegraphics[width=\columnwidth]{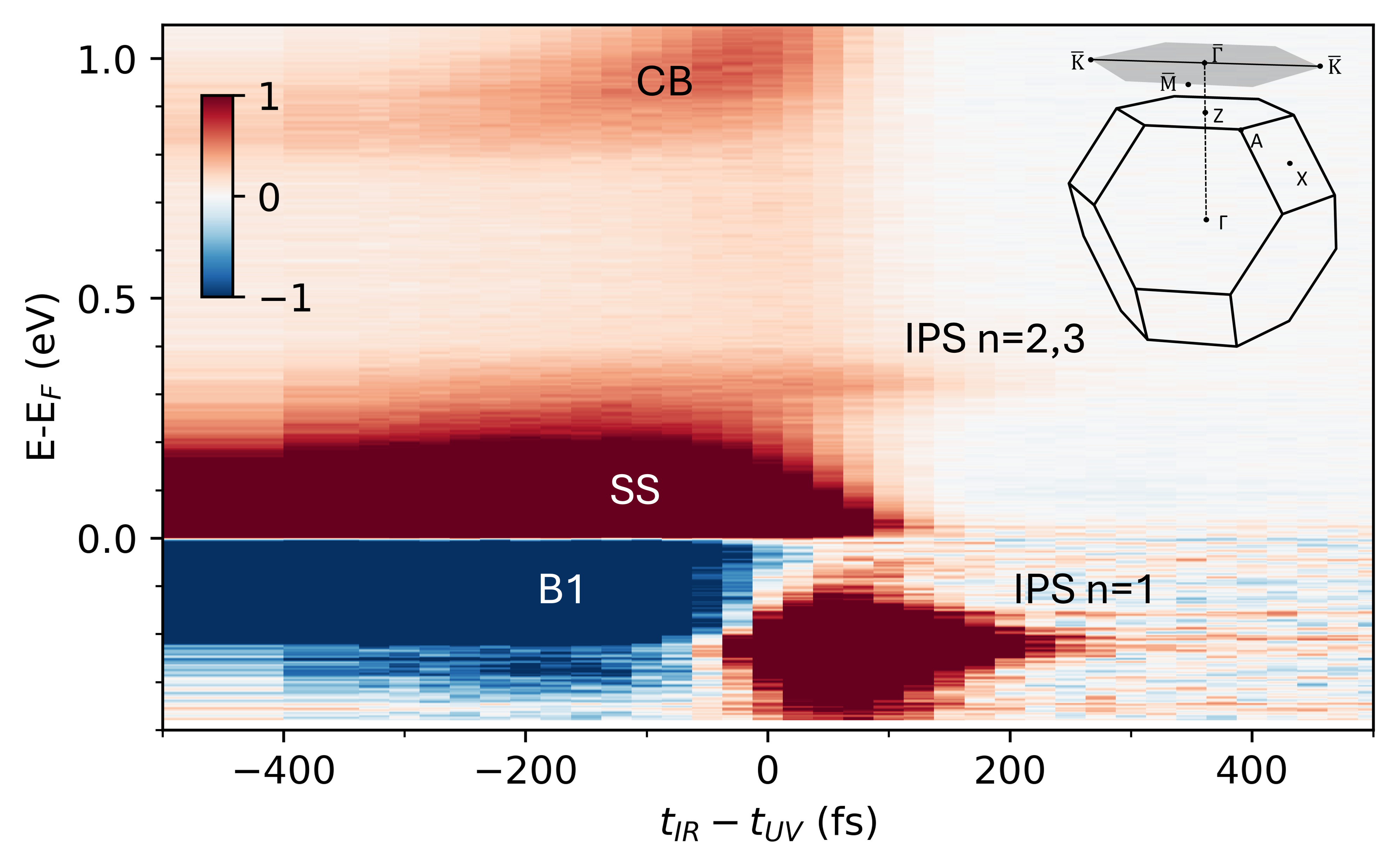}
   \end{subfigure}\end{center}
    \caption{\textbf{Time-Resolved evolution of the Bandstructure of $\alpha$-GeTe(111) :} Difference map of the intensity at normal emission between every delay and the one at $t_{IR}-t_{UV}=650$ fs, assumed to be close to equilibrium. Red and blue colors indicate an increase or loss in spectral weight with respect to equilibrium, respectively. Inset : Bulk Brillouin zone of GeTe and its surface projected plane along the [111] direction.}
    \label{fig:1-2}
\end{figure}

The electronic bandstructure of $\alpha-$GeTe(111) is presented in Fig. \ref{fig:1a}, in which we measured with a probe photon energy $h\nu=21.2$ eV the valence band (VB) structure along the high-symmetry line $\overline{K\Gamma K}$ (see black line in Brillouin zone in inset of Fig. \ref{fig:1-2}). 
Several electronic states are observed in close proximity to the Fermi level. Notably, an intense surface state (SS) is detected between 0.3 and 0.4 \AA$^{-1}$. Additionally, near normal emission, a bulk state (B1) is identified, in addition to a surface resonance state (SR). The SR state is characterized by a wavefunction originating from the bulk, with high amplitude at the surface \cite{krempasky_disentangling_2016,han_electronic_2014}. Different others bulks states can be identified from 1 to 3 eV below the Fermi level.

The states close to $E_F$ exhibit all Rashba-type spin splitting, resulting from inversion symmetry breaking. For the SS, this occurs at the sample surface, while for B1, it is attributed to the ferroelectric distortion within the bulk \cite{rashba_symmetry_nodate,bychkov1984properties,Rotenberg_Spin-Orbit_1999,chassot_persistence_2024}.

The measured electronic structure shows excellent agreement with previous reports \cite{di_sante_electric_2013,rinaldi_evidence_2016,liebmann_giant_2016,krempasky_entanglement_2016,elmers_spin_2016,kremer_unveiling_2020,ryu_chemical_2021,kremer_field-induced_2022}. Notably, the clear resolution of the splitting between the bulk state and the surface resonance state - often challenging to observe due to spectral broadening \cite{krempasky2020fully} - highlights the quality of the measurements.

Since static ARPES provides access only to occupied electronic states, we now turn to TR-ARPES. Using a UV photon energy of 6.33 eV, we measure the same electronic states as those observed at 21.2 eV, with the exception of a slight shift in the bulk band B1 due to dispersion along $k_\perp$ (see Fig. \ref{fig:1b}, scanning the red region highlighted in Fig. \ref{fig:1a}). By photoexciting the material with an IR pulse at 1.55 eV, we transiently populate the conduction band (CB). As illustrated in Fig. \ref{fig:1b}, this allows for precise measurement of the CB and SS above the Fermi level, in agreement with prior studies \cite{clark2022ultrafast,PhysRevB.111.235109}. We define here $t_0$ as the temporal overlap between the IR and UV pulses. Note that each time-resolved spectrum is normalized by the maximum of each momentum distribution curve (MDC), with some offset correction, to avoid enhancing a gap. This enables us to display in the same figure intense signals coming from occupied VB and the weaker response from electrons photoexcited in previously unoccupied states (e.g. CB or IPS).

\begin{figure*}
    \includegraphics[width=\linewidth]{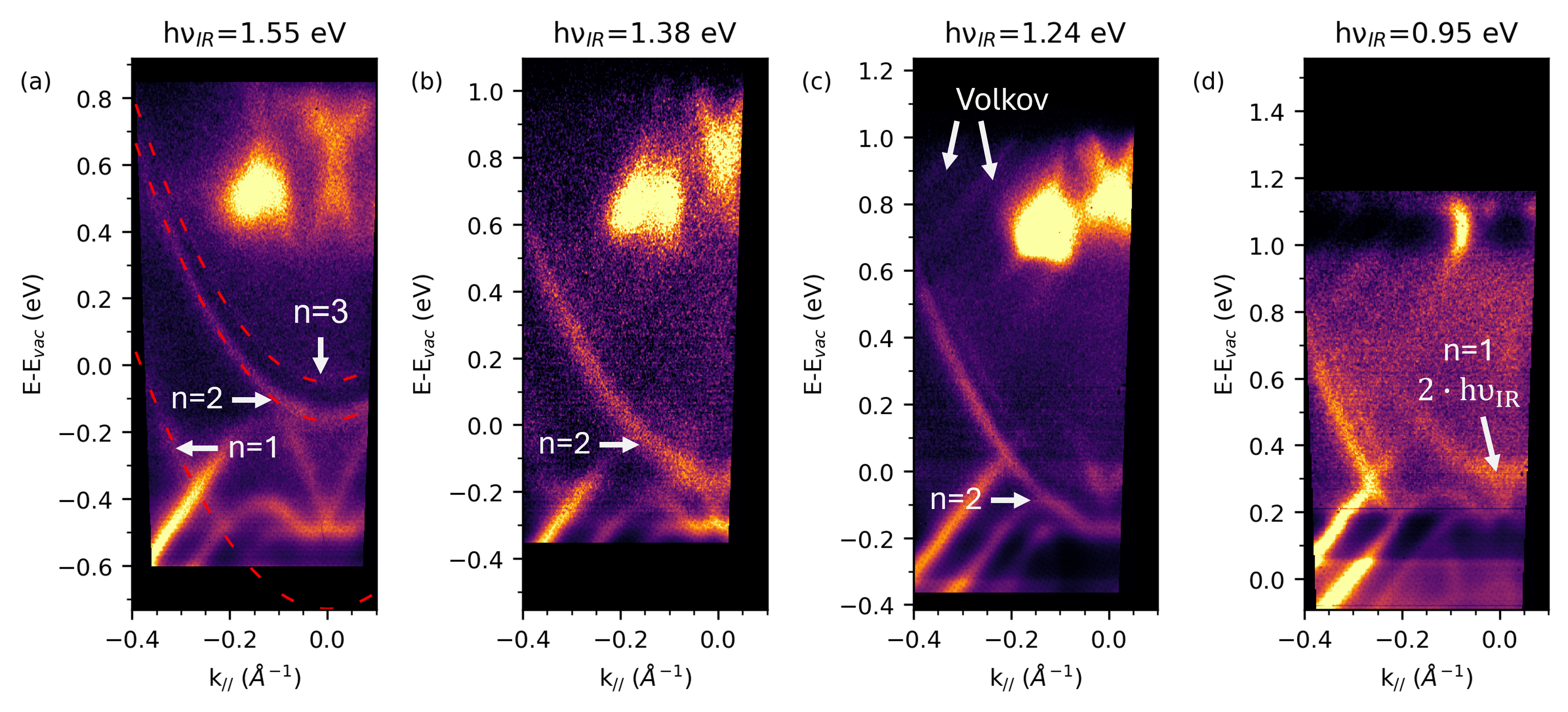}
    \begin{subfigure}[b]{0\linewidth}\phantomcaption\label{fig:2a}\end{subfigure}    \begin{subfigure}[b]{0\linewidth}\phantomcaption\label{fig:2b}\end{subfigure}
    \begin{subfigure}[b]{0\linewidth}\phantomcaption\label{fig:2c}\end{subfigure}
    \begin{subfigure}[b]{0\linewidth}\phantomcaption\label{fig:2d}\end{subfigure}
   \caption{\textbf{Evolution of the IPS as a function of IR photon energy.} Collection of time-resolved ARPES measurements along the high-symmetry line $\overline{K\Gamma}$, taken with an UV photon energy of 6.3 eV and for different IR energies as indicated at zero time delay. The IPS and replica are highlighted. The energy axis refers to the energy of the intermediate state in the two-photon transition with respect to the vacuum level. The energy axes are aligned such that the initial state energy is at constant height, as can be seen from the occupied surface states. The dashed line in panel (a) represents the parabolic curves used to extract the effective mass of the IPS.} 
    \label{fig:2}
\end{figure*}

The excitation dynamics are particularly evident in the difference plot of Fig. \ref{fig:1-2}, where the intensity at equilibrium is subtracted from the intensity at a specific time-delay (EDC integrated at normal emission over $\pm0.1$\AA$^{-1}$). In particular, by looking at the left part of the time-axis, i.e. \textbf{when the IR pulse comes before the UV pulse}, we can detect the creation of excited carriers: red regions indicate population of previously unoccupied states in the CB and SS above $E_F$, while blue regions represents a depopulation (hole creation) in the VB. Subsequently, photoexcited electrons relax their excess energy through electron-phonon scattering. Thereby, they minimize their energy and occupy lower electronic states within the same band, accumulating at the bottom of the CB before recombining with holes in the VB with a characteristic lifetime of $\approx 600$ fs. Additionally, shortly after photoexcitation, the electronic temperature rises significantly, resulting in a broadening of the Fermi-Dirac distribution (due to the intrinsic p-doping, the chemical potential lies close to the top of the VB and cut the SS). This accounts for the maximum population of the surface states around $\Delta t=-50$ fs, followed by a gradual relaxation as the energy is slowly transferred to the lattice  (see also our previous work \cite{PhysRevB.111.235109}).

Having now clearly established the occupied and unoccupied bandstructure, we now turn our attention to the positive part of the time-axis in Fig. \ref{fig:1-2}, i.e. \textbf{when the UV comes before the IR pulses}. For an IR beam linearly polarized along the direction parallel to the plane of incidence, three parabolas appear as shown in Fig. \ref{fig:1c}. 
By looking at their time-dynamics (see Fig. \ref{fig:1-2}) and specifically their relaxations towards positive IR-UV delays, one observes that those states are photoexcited by the UV pulse and photoemitted by the IR pulse, in contrary to the dynamics of the CB for negative IR-UV delays.

Combined with the fact that the isotropic parabolas, absent under s-polarized IR excitation (not shown), follow a quasi-free electron dispersion with an effective mass around $0.7\cdot m_e$ (see more details in Table \ref{tabel}), this strongly supports their interpretation as UV-populated image-potential states.
We note that the effective mass of IPS is usually expected to be around 1, although different masses have already been observed \cite{giesen1987effective,ferrini2003effective,rohleder_momentum-resolved_2005,rohleder_time-resolved_2005,ponzoni2023dirac}. We hypothesize that the semiconducting nature of p-doped GeTe surface weakens the dielectric response, thereby decreasing the Coulomb interaction between the IPS electron and the image charge. This intuitively delocalizes the IPS in real space, therefore decreasing its effective mass.

To enable a more rigorous analysis of the data, particular attention must be paid to the relevant energy scales. In conventional photoemission spectroscopy, the final state energy of electrons is expressed with respect to the Fermi level, in order to retrieve their binding energy. This is however not the case for IPS, which are bound states relative to the vacuum level:
Thus, from now a different energy scale will be used, namely $E-E_{\text{Vac}}$, i.e. the energy of the intermediate state in the two-photon transition with respect to the vacuum level (see appendix for an explanation of the conversion).

\begin{figure*} 
    \includegraphics[width=\linewidth]{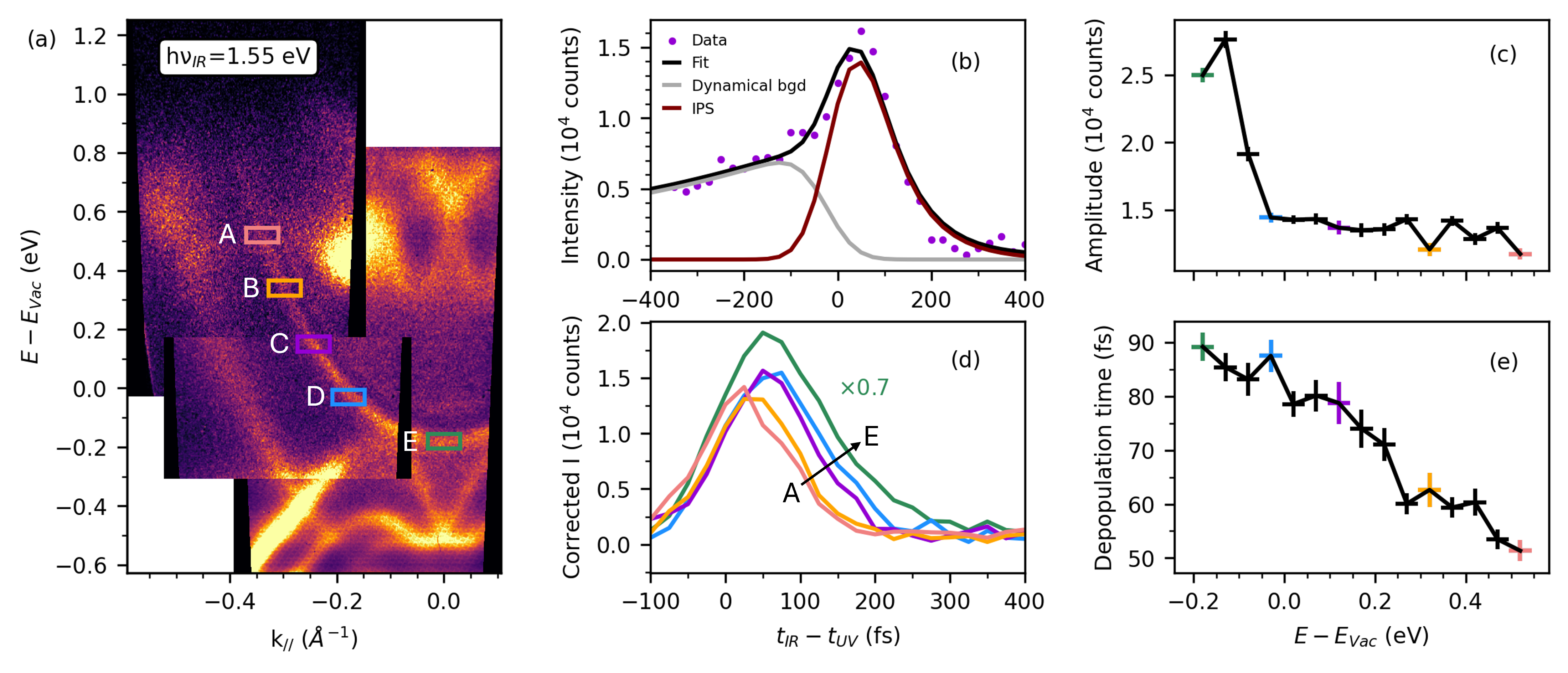}
    \begin{subfigure}[b]{0\linewidth}\phantomcaption\label{fig:3a}\end{subfigure}    \begin{subfigure}[b]{0\linewidth}\phantomcaption\label{fig:3b}\end{subfigure}
    \begin{subfigure}[b]{0\linewidth}\phantomcaption\label{fig:3c}\end{subfigure}
    \begin{subfigure}[b]{0\linewidth}\phantomcaption\label{fig:3d}\end{subfigure}
    \begin{subfigure}[b]{0\linewidth}\phantomcaption\label{fig:3e}\end{subfigure}
   \caption{\textbf{Evolution of the IPS below and above the vacuum level.} (a) Collection of time-resolved ARPES measurements along the high-symmetry line $\overline{K\Gamma}$, taken with an UV photon of energy 6.3 eV and an IR photon of energy 1.55 eV that arrive simultaneously at $t_0$ to follow the whole dispersion of the $n=2$ and $n=3$ IPS above and below the vacuum level. (b) Exemplary time trace of the $n=2$ IPS (at $E-E_{Vac}=0.12$ eV - see purple box C in (a)) as a function of IR-UV delay with a fitting procedure (black) to decompose the dynamical background (gray) and the IPS contribution from the signal (dark red). (c) Evolution of the fitted intensity of the $n=2$ IPS as a function of its energy to vacuum level. (d) Collection of Intensity time traces of the $n=2$ at different $E-E_{\text{Vac}}$ (see boxes in (a) labelled from A through E) after dynamical background subtraction. Note that the green curve has been multiplied by a factor of $0.7$ for the sake of comparison. (e) Evolution of the fitted depopulation time of the $n=2$ IPS as a function of energy to the vacuum level.}
    \label{fig:3}
\end{figure*}

Using this new energy axis, we therefore plot the photoemission data at time delay set to $t_0$ in Fig. \ref{fig:1c} to Fig. \ref{fig:2a}, to focus only on the free-electron-like parabolic dispersions. Two such features that are detected above the surface states and one below (see white arrows and dashed curves in Fig. \ref{fig:2a}). Further confirmation of the occurrence of IPS is coming from their dispersion upon varying the IR energy, as the final state energy changes linearly with slope 1, as shown in Figs \ref{fig:2a} to \ref{fig:2d}  where the parabola are kept at the same $E-E_{\text{Vac}}$ values. 

Note that the data in Fig. \ref{fig:2} have been taken with a the time-delay kept at zero (i.e. in coincidence of UV and IR pulses). This explains why the CB signal is stronger than in Fig. \ref{fig:1c} and why one can also detect weak replicas of the surface and valence bands, shifted by the energy of the pump photon (this is especially visible in Fig. \ref{fig:2c}), known as Volkov states or laser-assisted photoemission \cite{miaja2006laser}. Furthermore, in Fig. \ref{fig:2d}, an additional parabolic feature is observed. This is attributed to the occupation of the first IPS by one UV photon, followed by photoemission via absorption of \textit{two} IR photons, a process that accounts for the reduced intensity of this parabola.

\begin{figure}
    \includegraphics[width=0.95\linewidth]      {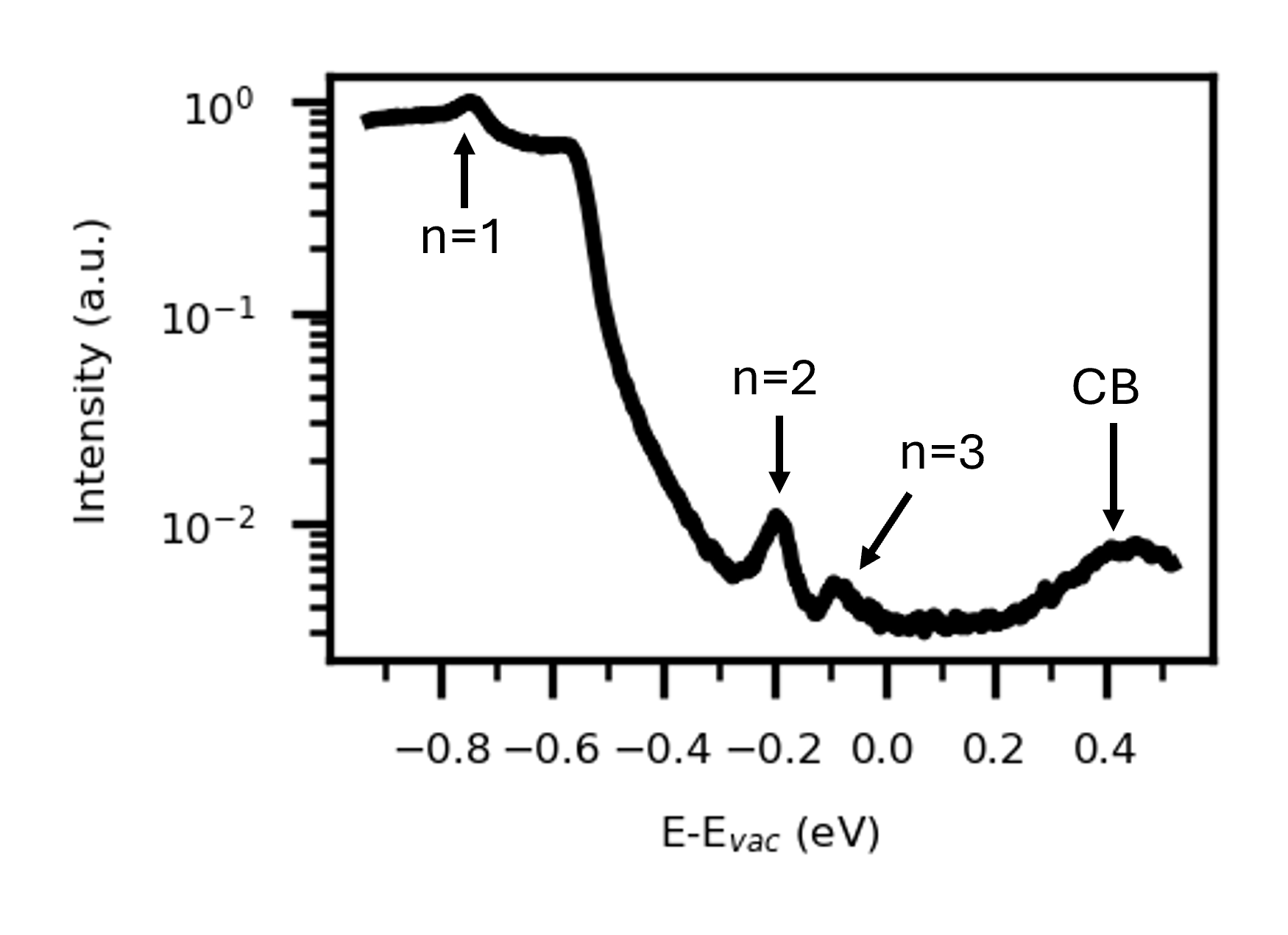}
    \caption{EDC at normal emission (integrated between $k_{\parallel}\in[-0.04,0.04]$ \AA$^{-1}$) of a TR-ARPES measurement along $\overline{K\Gamma K}$ with an UV photon energy of 6.33 eV and an IR photon energy of 1.55, shining simultaneously on the sample with highlighted position of the IPS.} 
    \label{figAppendix:EDC}
\end{figure}

To further refine the determination of binding energies, EDCs are extracted at normal emission (see Fig. \ref{figAppendix:EDC}). This analysis enables the characterization of the binding energies of the first three IPS, as summarized in Table \ref{tabel}. Those clearly follow the expected Rydberg series, consistent with theoretical predictions expressed in the Hydrogen-like equation:\\
$E_B(n)=\frac{\epsilon-1}{\epsilon+1}\cdot\frac{R_y}{16}\cdot \frac{1}{(n+a)^2} =\frac{\epsilon-1}{\epsilon+1}\cdot\frac{0.85 eV}{(n+a)^2} $, with a quantum defect parameter that we estimate to be $a\approx0.05\pm0.04$ and a dielectric constant of GeTe of $\epsilon=30.5\pm2.2$, close to literature values \cite{tsu1968optical,chen2017dielectric}.

\begin{table}[H]
    \centering
    \begin{tabular}{|c|c|c|c|}
        \hline
         & $n=1$ & $n=2$  & $n=3$\\
         \hline
        $E_B(n)$ & $0.719 \pm0.010$ eV & $0.193\pm0.010$ eV & $0.079\pm0.010$ eV \\
        \hline        $m_\text{eff}/m_\text{e}$&$0.76\pm0.03$&$0.72\pm0.03$&$0.72\pm0.03$\\
        \hline
    \end{tabular}
    \caption{Binding energy and effective masses for the different IPS in GeTe(111). Note that the experimental uncertainties are here dominated by our capacity to extract accurately the workfunction. The effective masses are extracted by fitting the IPS parabola (see e.g. Fig. \ref{fig:2a}).}
    \label{tabel}
\end{table} 

Given the clear identification of the IPS, we trace their full dispersion until they disappear at high energy, as shown in Fig. \ref{fig:3a}. This allows us to unambiguously demonstrate their remarkable and highly unexpected extension reaching up to 0.8 eV above the vacuum level. Such a complete dispersion, significantly exceeding the vacuum level, has not been previously reported in the literature. While IPS detection is still relatively common with TR-ARPES (see e.g. Ref. \cite{sobota_ultrafast_2012}), prior observations have been limited to parabolic dispersions terminating at the vacuum level - a representative example being shown in the prototypical sample Bi$_2$Se$_3$ in the appendix (Fig. \ref{figAppendix:Bi2Se3}). 
Notably, in our case, this extension appears to be independent of both the IPS quantum number (the expansion above $E_{\text{Vac}}$ being similar for the three observed IPS) and the infrared excitation energy. For instance, the dispersion of the IPS can be tracked up to at least 0.75 eV above the vacuum level for an infrared photon energy of $h\nu_{IR}=0.95$ eV (see Fig. \ref{fig:2d}).

To go deeper in the analysis of the evolution of the IPS above and below the vacuum level, we now focus on the second IPS, which is well separated from the occupied states and has high intensity. To investigate its evolution along the full dispersion, we extract the intensity as a function of IR-UV pulse delay within energy-momentum regions spaced by 50 meV in binding energy, each with a width of 50 meV and a momentum range of 0.06 \AA$^{-1}$ (see a selection of the colored boxes in Fig. \ref{fig:3a}). A careful look at their dynamics in Fig. \ref{fig:3b} reveals the following phenomena: 

At negative IR–UV delays (IR preceding UV), the photoemission intensity decays exponentially, consistent with conventional TR-ARPES excitation. In this regime, electrons are promoted to unoccupied CB states and relax to equilibrium within 4 ps, as reported previously \cite{PhysRevB.111.235109}. During photoemission, some carriers undergo inelastic scattering, producing a diffuse background across the detector. This dynamic background spans energies below the conduction band, including the band gap, and is evident in Fig. \ref{fig:1-2} as an increased signal (red) in the range $E-E_F\in[0.4,0.8]$ eV. Its temporal evolution follows an exponential decay with a lifetime comparable to that of the carriers. We model this decay using an exponential function (gray curve in Fig. \ref{fig:3b}).

Second, at positive IR-UV delays, an additional, shorter but more intense exponential decay is observed, corresponding to the depopulation of the IPS. The combination of these two processes is modeled using a double-exponential decay convolved with a Gaussian function ($\sigma=45$ fs) representing the pump-probe cross-correlation, to isolate the IPS dynamics. In Fig. \ref{fig:3d}, we present the signal extracted from the boxes indicated in Fig. \ref{fig:3a}, after subtracting the dynamic background obtained from the fit described above. This reveals that the signal extracted near the top of the IPS dispersion (see salmon-colored curve) reaches its maximum close to $t_0$, with a slight delay attributed to the presence of an intermediate state in the photoemission process \cite{hertel_ultrafast_1996}. Furthermore, the position of the maximum shifts toward longer IR–UV delays for lower energies. This is consistent with intraband scattering, for which electrons undergo momentum relaxation as they move toward the dispersion minimum \cite{berthold_momentum-resolved_2002}.

To further quantify these observations, we examine the fitted amplitude and depopulation time of the IPS in Figs. \ref{fig:3c} and \ref{fig:3e}. Two key trends emerge: (i) the signal amplitude decreases significantly near the vacuum level but remains approximately constant above it (Fig. \ref{fig:3c}); and (ii) the IPS lifetime decreases linearly with increasing energy, as shown in Fig. \ref{fig:3e} and directly reflected in the temporal traces of Fig. \ref{fig:3c}.

\section{\label{sec:Discussion}Discussion}

\begin{figure}\begin{center}        \includegraphics[width=\linewidth]{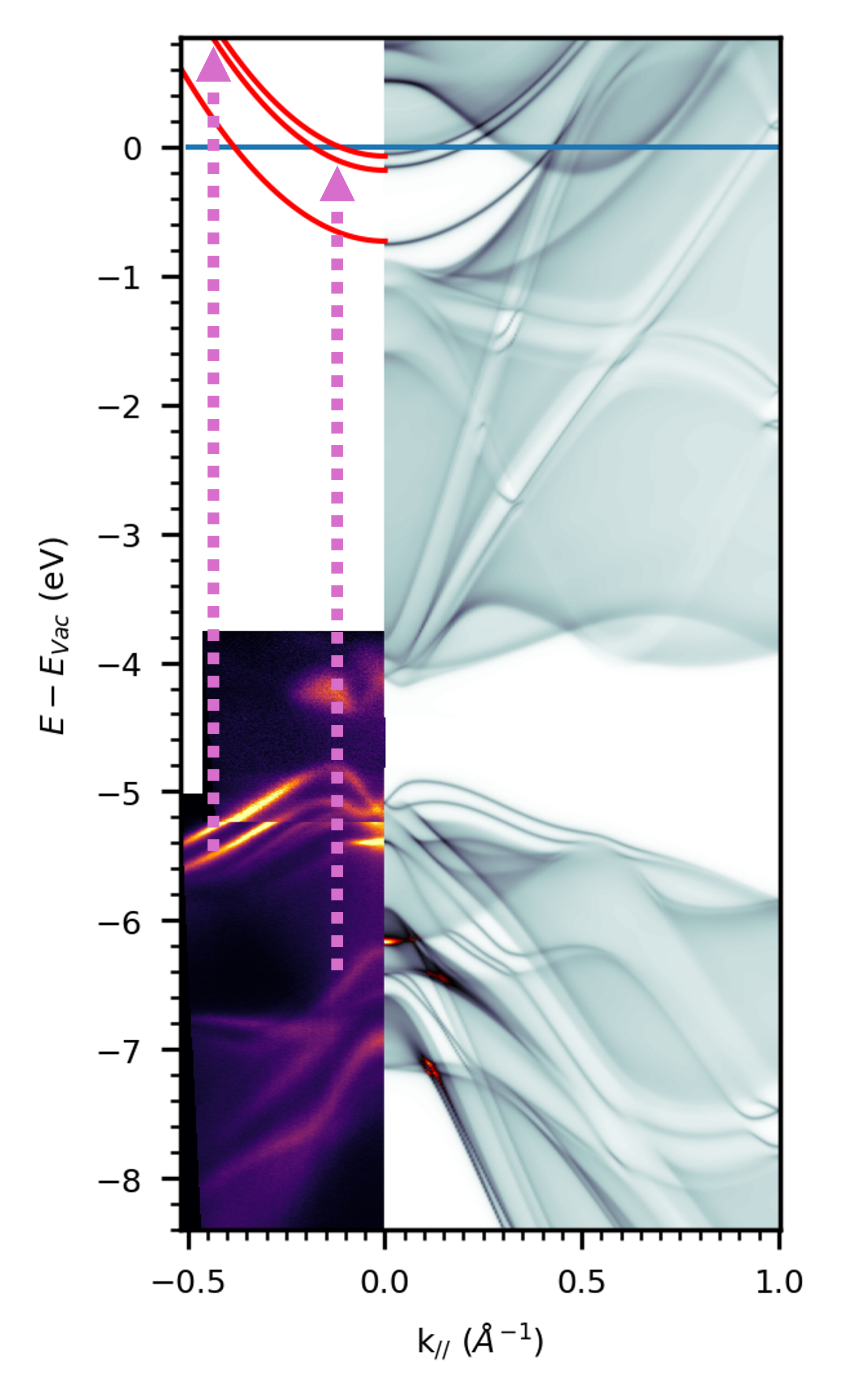}
\caption{Comparison between the experimentally observed bandstructure (same parameters as in Fig. \ref{fig:1}), theoretical calculations, and projected IPS. Pink arrows indicate the different reservoirs and population pathways feeding the intermediate IPS states.} 
\label{fig:4}
\end{center}\end{figure}
To develop an explanation for the presence of IPS far above the vacuum level, it is first essential to obtain a clear picture of the initial and intermediate states inside the crystal. To this end, Fig. \ref{fig:4} reconstructs the initial electronic structure, illustrating the VB and CB based on our measurements in Fig. \ref{fig:1}, projecting the IPS positions as intermediate states in the 2-photon photoemission process. This experimental projection of the IPS within the crystal is represented by the red parabolas in Fig. \ref{fig:4}. For comparison, we also provide a theoretical prediction of the electronic structure by computing the ground-state Bloch spectral function using Green’s function formalism. Additional details are given in Sec. \ref{Section_Method}. These fully self-consistent calculations for a semi-infinite surface were modified to include an analytical continuation of the surface barrier with a $1/z$ convergence toward the vacuum level, rather than the exponential convergence to the work function typical of conventional DFT. By introducing a sufficiently large vacuum region, we can model the electronic structure of surface states, bulk states (each layer of the semi-infinite surface contributing to the dark-shaded bulk continuum due to $k_\perp$ dispersion), and most importantly, the IPS dispersions. In addition, to correct for the underestimated gap within LSDA, the gap between the VB and the CB has been artificially increased by 0.38 eV.
In the Appendix (see Fig. \ref{figA:Extension}), we further decompose the calculated electronic density layer-by-layer to confirm that these states originate from the first few layers extending into the vacuum. The results of these calculations are shown on the right side of Fig. \ref{fig:4}. Beyond the excellent agreement with experiment, these results clarify certain spectroscopic signatures: first, the bottoms of the IPS dispersions are located in a band gap of the material, a necessary condition for the existence of a series of discrete states (otherwise, they merge into the continuum of bulk states). Second, focusing on the $n=1$ IPS, the calculations reveal that it merges into the bulk continuum at $k = 0.4$ \AA$^{-1}$. This provides new scattering channels and explains why, in Fig. \ref{fig:3a}, the first IPS parabola appears much broader than those of $n = 2$ and $n = 3$ for high momentum values.

Building on this understanding and on the fact that the measured extension of IPS above the vacuum level does not depend on the IR photon energy, we propose the following scenario, based on the interplay of two effects specific to GeTe. First, the system benefits from large initial-states reservoirs, either from surface states present at $k > 0.3$ \AA$^{-1}$ or from the bulk continuum located 1 eV below the Fermi level at $\overline{\Gamma}$ (see the two excitation pathways for 6.3 eV photons indicated by pink arrows in Fig. \ref{fig:4}). The presence and contribution of these reservoirs likely influence the amplitude of the IPS as a function of $E-E_{\text{Vac}}$ 
(see Fig. \ref{fig:3c}). Second, a strong dipole transition between initial and intermediate states is expected, owing to the polar nature of ferroelectric GeTe. 

These factors can account for the strong Rabi oscillations between initial states and IPS under the UV pulse electric field, leading to significant population transfer. This coherent superposition persists as long as the photon field is present. Within the decoherence time, electrons may remain with a non-zero probability in these excited states - even above the vacuum level - long enough to be detected, explaining why the depopulation time does not fall to zero above the vacuum level in the case of GeTe (see Fig. \ref{fig:3e}). Finally, our significant IR fluence favors the photoemission of IPS.

\section{\label{sec:Conclusion}Conclusion}

In conclusion, we have unambiguously demonstrated the presence of image-potential states in GeTe. By combining high-resolution TR-ARPES measurements with a novel computational approach based on Bloch spectral functions, we achieve an exceptional level of agreement between experiment and theory. Remarkably, we observe that three IPS extend up to 0.8 eV above the vacuum level. We propose that it arises from the strong dipolar response with the considerable availability of initial states whose energy separation from the intermediate IPS matches the UV excitation. This condition allows electrons to remain above the vacuum level within the decoherence time, provided the detection occurs sufficiently fast.

We anticipate that these findings will stimulate further theoretical and experimental efforts. Future studies could explore the role of ferroelectricity in IPS formation, investigate dephasing dynamics with improved temporal resolution, and employ spin-resolved ARPES or CD-2PPE to confirm the Rashba splitting predicted for these states (see Fig. \ref{figA:Rashba}). 

\section{\label{sec:Appendix}Appendix}

\textbf{Conversion of the energy axis :} 
In photoemission spectroscopy, we measure the kinetic energy of photoelectrons $E_\text{kin}$. It is well known how to retrieve information on the binding energy from the final states originating from the VB or the CB, as they are photoemitted from the absorption of an UV photon of energy $h\nu$ ($E_B^{VB}=h\nu-\phi-E_\text{kin}$, where $\phi$ is the workfunction).
However, the situation differs for intermediate states such as IPS, as they are populated via UV excitation and subsequently photoemitted through IR photon absorption. In this case, the binding energy of the IPS is defined with respect to the vacuum level.

Practically, this conversion is achieved through the difference $\Delta E$ of the measured kinetic energy of the IPS final state with that of a reference state, such as the Fermi level (i.e. our energy axis in Fig. \ref{fig:1}). The initial binding energy of the IPS relative to the vacuum level can then be calculated using the following expression :
\begin{equation}
    \begin{split}
    \Delta E&:=E^\text{kin}_{IPS}-E^\text{kin}_{F}\\
    &=(h\nu_{IR}-E_B(IPS))-(h\nu_{UV}-\phi)\\
    &\rightarrow E_B(IPS)= h\nu_{IR}-h\nu_{UV}+\phi-\Delta E
    \end{split}
\end{equation}

For consistency, to describe the energy of the intermediate state with respect to the vacuum level, we use $E-E_{vac}=-E_{B}$ as the energy axis for Figs \ref{fig:2} to \ref{fig:4}.

\begin{figure}\begin{center}
\includegraphics[width=\linewidth]      {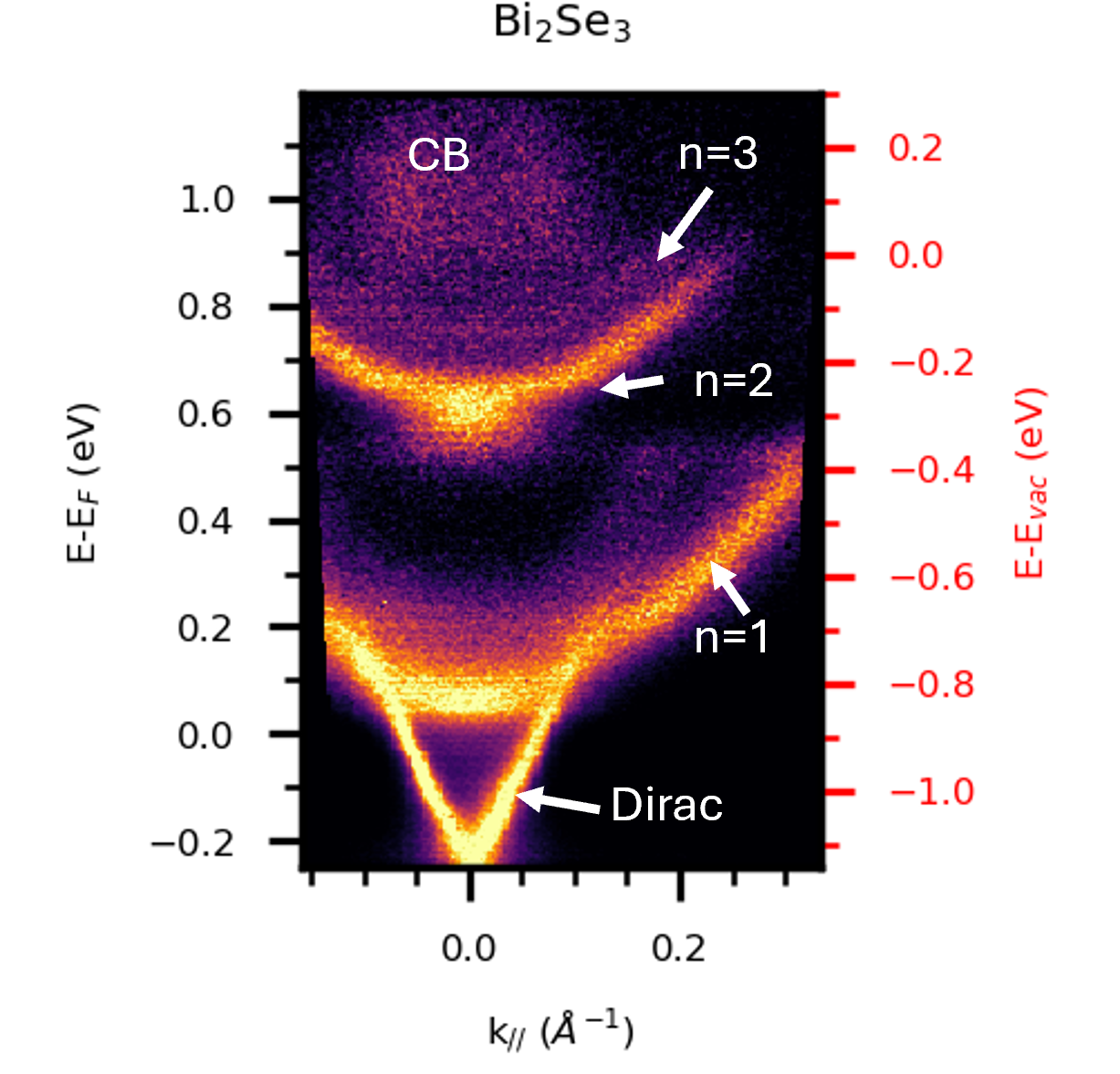}
\caption{TR-ARPES spectrum of Bi$_2$Se$_3$ at 80 K taken with UV and IR photon energies of respectively 6.3 eV and 1.55 eV, integrated between a IR-UV delay of $t_0$ and $t_0+300$ fs.} 
\label{figAppendix:Bi2Se3}
\end{center}\end{figure}

\begin{figure}\begin{center}     \includegraphics[width=\linewidth]{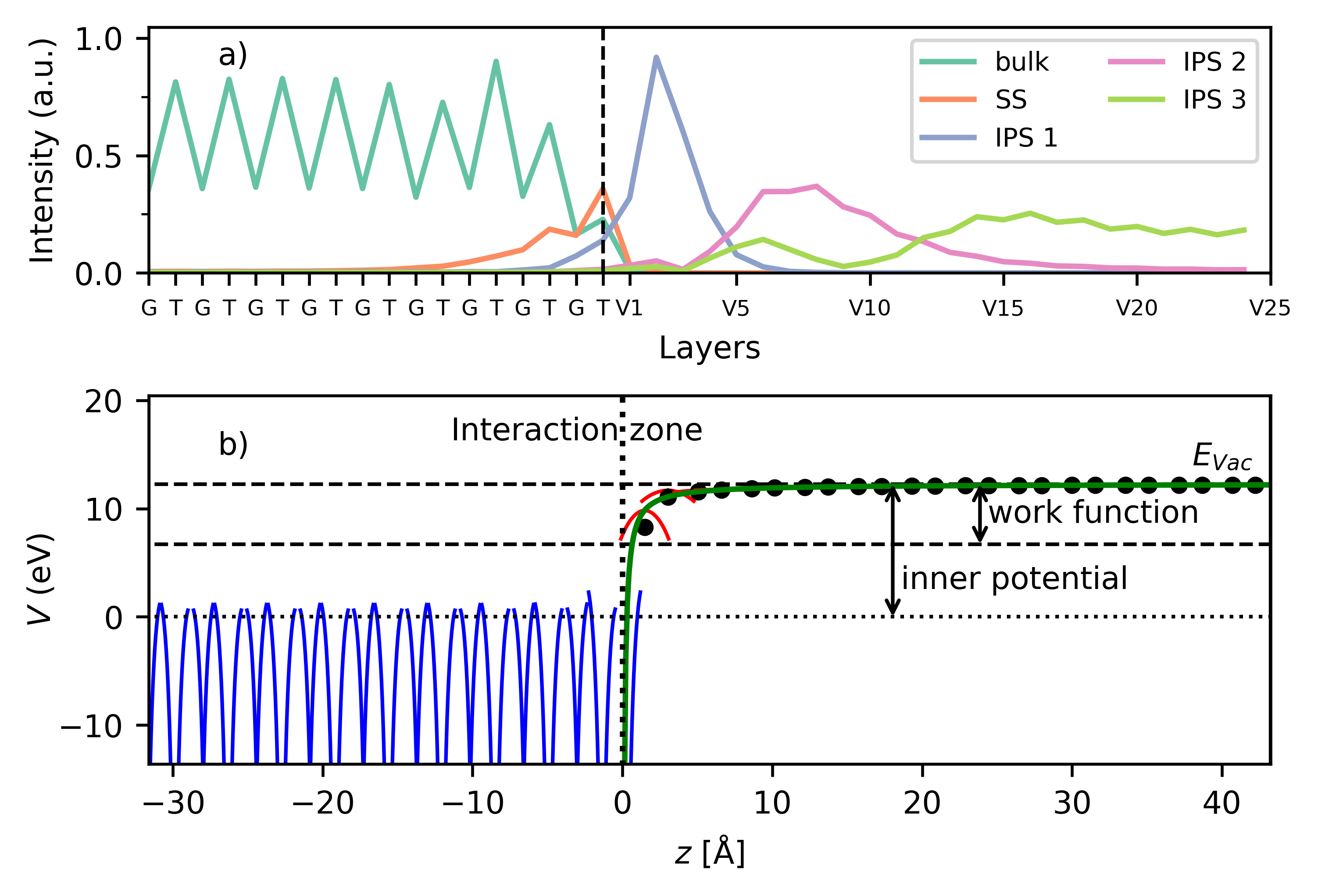}
\caption{(a) Layer-resolved intensity of the Bloch spectral function integrated within a small energy–momentum window, plotted as a function of layer index in the semi-infinite crystal and vacuum region (G stands for Ge, T for Te, V1 for the first vacuum layer, ...). The integration windows are selected to isolate the specific states indicated in the legend ; (b) Evolution of the potential in the interaction zone with in blue the atomic potential, black the vacuum layer, with respective potential in red and in green the $1/z$ convergence of the surface barrier to the vacuum level. Note that the x-axis for panel (a) and (b) are consistent.} 
\label{figA:Extension}
\end{center}\end{figure}

\textbf{Layer-resolved contributions:} As discussed in the main text, each layer of the semi-infinite surface contributes to the calculation of the total electronic structure. However, it is possible to separate these contributions by integrating the intensity of the Bloch spectral function and, instead of summing over all layers, examining the spatial origin of the spectral weight. By carefully selecting the integration region to isolate a specific state - whether bulk, surface, or image potential states - we can gain insight into the spatial origin of these states.
Our calculations clearly demonstrate that a sufficiently large vacuum region is required to accurately capture the IPS. In particular, as shown in Fig. \ref{figA:Extension}, the full spatial extent of the third IPS is not even entirely contained within the computational domain. Note that, while in systems with an exponential potential, such as 2DEG, the states are localized just a few \AA{} from the surface \cite{moser2018extract}, in our case the vacuum potential barrier follows a $1/z$ shape, which explains why IPS extend further into the vacuum.

\begin{figure}\begin{center}        \includegraphics[width=0.75\linewidth]{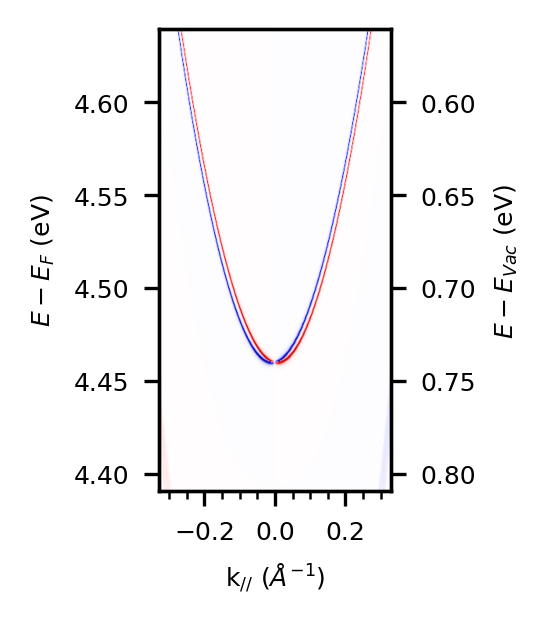}
\caption{Spin texture of the first IPS obtained through Bloch spectral function calculation.} 
\label{figA:Rashba}
\end{center}\end{figure}

From the results of our calculations, we can extract the spin-resolved structure of IPS, with a Rashba splitting of the order 8 meV, as shown in Fig. \ref{figA:Rashba}. To verify this, we suggest to use 2PPE in combination with circular dichroism, as it has been shown to be sensitive the small Rashba splitting in IPS states \cite{braun2021impact,tognolini2015rashba,nakazawa2016rashba,schottke2022rashba}. 

\section*{Acknowledgements}
G.S. would like to thank the Austrian Science Fund (FWF), who supported this study with projects PIN6540324.

J.M. and A.P. thank the project Quantum Materials for applications in sustainable technologies (QM4ST), funded as Project No. CZ.02.01.01/00/22{\_}008/0004572 by Programme Johannes Amos Comenius, call Excellent Research.

Skillful technical assistance was provided by F. Bourqui, B. Hediger and M. Andrey.
\newpage
\bibliography{MyLibrary}

@article{rashba_symmetry_nodate,
	title = {Symmetry of {Energy} {Bands} in {Crystals} of {Wurtzite} {Type} {II :} {Symmetry} of {Bands} with {Spin}-{Orbit} {Interaction} {Included}},
	author = {Rashba, E I and Sheka, V I},
	pages = {16},
	journal = {Fiz. Tverd. Tela},
	year = {1958},
}

@article{clark2022ultrafast,
  title={Ultrafast thermalization pathways of excited bulk and surface states in the ferroelectric rashba semiconductor GeTe},
  author={Clark, Oliver J and Wadgaonkar, Indrajit and Freyse, Friedrich and Springholz, Gunther and Battiato, Marco and S{\'a}nchez-Barriga, Jaime},
  journal={Advanced Materials},
  volume={34},
  number={24},
  pages={2200323},
  year={2022},
  publisher={Wiley Online Library}
}

@article{kremer_unveiling_2020,
	title = {Unveiling the complete dispersion of the giant {Rashba} split surface states of ferroelectric {$\alpha$}-{GeTe}(111) by alkali doping},
	volume = {2},
	url = {https://link.aps.org/doi/10.1103/PhysRevResearch.2.033115},
	doi = {10.1103/PhysRevResearch.2.033115},
	number = {3},
	urldate = {2022-03-30},
	journal = {Phys. Rev. Research},
	author = {Kremer, G. and Jaouen, T. and Salzmann, B. and Nicolaï, L. and Rumo, M. and Nicholson, C. W. and Hildebrand, B. and Dil, J. H. and Minár, J. and Springholz, G. and Krempaský, J. and Monney, C.},
	month = jul,
	year = {2020},
	pages = {033115},
}

@article{picozzi_ferroelectric_2014,
	title = {Ferroelectric {Rashba} semiconductors as a novel class of multifunctional materials},
	volume = {2},
	issn = {2296-424X},
	url = {https://www.frontiersin.org/article/10.3389/fphy.2014.00010},
	urldate = {2022-03-30},
	journal = {Frontiers in Physics},
	author = {Picozzi, Silvia},
	year = {2014},
}

@article{di_sante_electric_2013,
	title = {Electric {Control} of the {Giant} {Rashba} {Effect} in {Bulk} {GeTe}},
	volume = {25},
	issn = {1521-4095},
	url = {https://onlinelibrary.wiley.com/doi/abs/10.1002/adma.201203199},
	doi = {10.1002/adma.201203199},
	number = {4},
	urldate = {2022-03-30},
	journal = {Advanced Materials},
	author = {Di Sante, Domenico and Barone, Paolo and Bertacco, Riccardo and Picozzi, Silvia},
	year = {2013},
	pages = {509--513},
}

@article{ryu_chemical_2021,
	title = {Chemical control of the {Rashba} spin splitting size of {$\alpha$}-{GeTe}(111) surface states by adjusting the potential at the topmost atomic layer},
	volume = {103},
	url = {https://link.aps.org/doi/10.1103/PhysRevB.103.245113},
	doi = {10.1103/PhysRevB.103.245113},
	number = {24},
	urldate = {2022-03-30},
	journal = {Phys. Rev. B},
	author = {Ryu, Hanyoung and Lihm, Jae-Mo and Cha, Joonil and Kim, Beomyoung and Kim, Beom Seo and Kyung, Wonshik and Song, Inkyung and Kim, Yeongkwan and Han, Garam and Denlinger, Jonathan and Chung, In and Park, Cheol-Hwan and Park, Seung Ryong and Kim, Changyoung},
	month = jun,
	year = {2021},
	pages = {245113},
}

@article{liebmann_giant_2016,
	title = {Giant {Rashba}-{Type} {Spin} {Splitting} in {Ferroelectric} {GeTe}(111)},
	volume = {28},
	issn = {1521-4095},
	url = {https://onlinelibrary.wiley.com/doi/abs/10.1002/adma.201503459},
	doi = {10.1002/adma.201503459},
	number = {3},
	urldate = {2022-03-30},
	journal = {Advanced Materials},
	author = {Liebmann, Marcus and Rinaldi, Christian and Di Sante, Domenico and Kellner, Jens and Pauly, Christian and Wang, Rui Ning and Boschker, Jos Emiel and Giussani, Alessandro and Bertoli, Stefano and Cantoni, Matteo and Baldrati, Lorenzo and Asa, Marco and Vobornik, Ivana and Panaccione, Giancarlo and Marchenko, Dmitry and Sánchez-Barriga, Jaime and Rader, Oliver and Calarco, Raffaella and Picozzi, Silvia and Bertacco, Riccardo and Morgenstern, Markus},
	year = {2016},
	        
	pages = {560--565},
}

@article{rinaldi_ferroelectric_2018,
	title = {Ferroelectric {Control} of the {Spin} {Texture} in {GeTe}},
	volume = {18},
	issn = {1530-6984},
	url = {https://doi.org/10.1021/acs.nanolett.7b04829},
	doi = {10.1021/acs.nanolett.7b04829},
	number = {5},
	urldate = {2022-03-30},
	journal = {Nano Lett.},
	author = {Rinaldi, Christian and Varotto, Sara and Asa, Marco and Sławińska, Jagoda and Fujii, Jun and Vinai, Giovanni and Cecchi, Stefano and Di Sante, Domenico and Calarco, Raffaella and Vobornik, Ivana and Panaccione, Giancarlo and Picozzi, Silvia and Bertacco, Riccardo},
	month = may,
	year = {2018},
	pages = {2751--2758},
}

@article{krempasky_spin-resolved_2019,
	series = {Spin-{Orbit} {Coupled} {Materials}},
	title = {Spin-resolved electronic structure of ferroelectric {$\alpha$}-{GeTe} and multiferroic {Ge1}-{xMnxTe}},
	volume = {128},
	issn = {0022-3697},
	url = {https://www.sciencedirect.com/science/article/pii/S0022369717314531},
	doi = {10.1016/j.jpcs.2017.11.010},
	
	urldate = {2022-03-30},
	journal = {Journal of Physics and Chemistry of Solids},
	author = {Krempaský, J. and Fanciulli, M. and Pilet, N. and Minár, J. and Khan, W. and Muntwiler, M. and Bertran, F. and Muff, S. and Weber, A. P. and Strocov, V. N. and Volobuiev, V. V. and Springholz, G. and Dil, J. H.},
	month = may,
	year = {2019},	
	      
	pages = {237--244},
}

@article{krempasky_entanglement_2016,
	title = {Entanglement and manipulation of the magnetic and spin–orbit order in multiferroic Rashba semiconductors},
	volume = {7},
	issn = {2041-1723},
	url = {https://www.nature.com/articles/ncomms13071},
	doi = {10.1038/ncomms13071},
	
	number = {1},
	urldate = {2022-03-30},
	journal = {Nat Commun},
	author = {Krempaský, J. and Muff, S. and Bisti, F. and Fanciulli, M. and Volfová, H. and Weber, A. P. and Pilet, N. and Warnicke, P. and Ebert, H. and Braun, J. and Bertran, F. and Volobuev, V. V. and Minár, J. and Springholz, G. and Dil, J. H. and Strocov, V. N.},
	month = oct,
	year = {2016},
	pages = {13071},
}

@article{krempasky_disentangling_2016,
	title = {Disentangling bulk and surface Rashba effects in ferroelectric {$\alpha$}-{GeTe}},
	volume = {94},
	url = {https://link.aps.org/doi/10.1103/PhysRevB.94.205111},
	doi = {10.1103/PhysRevB.94.205111},
	pages = {205111},
	number = {20},
	year = {2016},
	journal = {Physical Review B},
	shortjournal = {Phys. Rev. B},
	author = {Krempaský, J. and Volfová, H. and Muff, S. and Pilet, N. and Landolt, G. and Radović, M. and Shi, M. and Kriegner, D. and Holý, V. and Braun, J. and Ebert, H. and Bisti, F. and Rogalev, V. A. and Strocov, V. N. and Springholz, G. and Minár, J. and Dil, J. H.},
	urldate = {2023-01-19},
	date = {2016-11-07},
}

@article{kremer_field-induced_2022,
	title = {Field-induced ultrafast modulation of Rashba coupling at room temperature in ferroelectric {$\alpha$}-{GeTe}(111)},
	volume = {13},
	year = {2022},
	issn = {2041-1723},
	url = {http://arxiv.org/abs/2204.11630},
	doi = {10.1038/s41467-022-33978-3},
	pages = {6396},
	number = {1},
	journal = {Nature Communications},
	shortjournal = {Nat Commun},
	author = {Kremer, Geoffroy and Maklar, Julian and Nicolaï, Laurent and Nicholson, Christopher W. and Yue, Changming and Silva, Caio and Werner, Philipp and Dil, J. Hugo and Krempaský, Juraj and Springholz, Gunther and Ernstorfer, Ralph and Minár, Jan and Rettig, Laurenz and Monney, Claude},
	urldate = {2023-01-19},
	date = {2022-10-27},
	eprinttype = {arxiv},
	eprint = {2204.11630 [cond-mat]},
}

@article{pawley_diatomic_1966,
	title = {Diatomic Ferroelectrics},
	volume = {17},
	url = {https://link.aps.org/doi/10.1103/PhysRevLett.17.753},
	doi = {10.1103/PhysRevLett.17.753},
	pages = {753--755},
	number = {14},
        year={1966},
	journal = {Physical Review Letters},
	shortjournal = {Phys. Rev. Lett.},
	author = {Pawley, G. S. and Cochran, W. and Cowley, R. A. and Dolling, G.},
	urldate = {2023-01-19},
	date = {1966-10-03},
}

@article{wang_spin_2020,
	title = {Spin Hall effect in prototype Rashba ferroelectrics {GeTe} and {SnTe}},
	volume = {6},
	rights = {2020 The Author(s)},
	issn = {2057-3960},
	url = {https://www.nature.com/articles/s41524-020-0274-0},
	doi = {10.1038/s41524-020-0274-0},
	pages = {1--7},
	number = {1},
	journal = {npj Computational Materials},
	shortjournal = {npj Comput Mater},
	author = {Wang, Haihang and Gopal, Priya and Picozzi, Silvia and Curtarolo, Stefano and Buongiorno Nardelli, Marco and Sławińska, Jagoda},
	urldate = {2023-01-19},
        year = {2020},
	date = {2020-01-24},
	langid = {english},
}

@article{Rotenberg_Spin-Orbit_1999,
  title = {Spin-Orbit Coupling Induced Surface Band Splitting in {Li/W}(110) and {Li/Mo}(110)},
  author = {Rotenberg, Eli and Chung, J. W. and Kevan, S. D.},
  journal = {Phys. Rev. Lett.},
  volume = {82},
  issue = {20},
  pages = {4066--4069},
  numpages = {0},
  year = {1999},
  month = {May},
  publisher = {American Physical Society},
  doi = {10.1103/PhysRevLett.82.4066},
  url = {https://link.aps.org/doi/10.1103/PhysRevLett.82.4066}
}

@article{bychkov1984properties,
  title={Properties of a 2D electron gas with lifted spectral degeneracy},
  author={Bychkov, Yua A and Rashba, {\'E} I},
  journal={JETP lett},
  volume={39},
  number={2},
  pages={78},
  year={1984},
}

@article{rinaldi_evidence_2016,
	title = {Evidence for spin to charge conversion in {GeTe}(111)},
	volume = {4},
	url = {https://aip.scitation.org/doi/10.1063/1.4941276},
	doi = {10.1063/1.4941276},
	pages = {032501},
	number = {3},
	journal = {{APL} Materials},
	author = {Rinaldi, C. and Rojas-Sánchez, J. C. and Wang, R. N. and Fu, Y. and Oyarzun, S. and Vila, L. and Bertoli, S. and Asa, M. and Baldrati, L. and Cantoni, M. and George, J.-M. and Calarco, R. and Fert, A. and Bertacco, R.},
	urldate = {2023-01-19},
	date = {2016-03},
        year = {2016},
}

@article{Ebert_2011,
doi = {10.1088/0034-4885/74/9/096501},
url = {https://dx.doi.org/10.1088/0034-4885/74/9/096501},
year = {2011},
month = {aug},
publisher = {},
volume = {74},
number = {9},
pages = {096501},
author = {H Ebert and D Ködderitzsch and J Minár},
title = {Calculating condensed matter properties using the {KKR}-{Green's} function method—recent developments and applications},
journal = {Reports on Progress in Physics}
}

@article{chassot_persistence_2024,
	title = {Persistence of Structural Distortion and Bulk Band Rashba Splitting in {SnTe} above Its Ferroelectric Critical Temperature},
	volume = {24},
	issn = {1530-6984},
	url = {https://doi.org/10.1021/acs.nanolett.3c03280},
	doi = {10.1021/acs.nanolett.3c03280},
	pages = {82--88},
	number = {1},
	journal = {Nano Letters},
	shortjournal = {Nano Lett.},
	author = {Chassot, Frédéric and Pulkkinen, Aki and Kremer, Geoffroy and Zakusylo, Tetiana and Krizman, Gauthier and Hajlaoui, Mahdi and Dil, J. Hugo and Krempaský, Juraj and Minár, Ján and Springholz, Gunther and Monney, Claude},
	urldate = {2025-03-07},
	date = {2024-01-10},
        year = {2024},
}

@article{han_electronic_2014,
	title = {Electronic structure of black phosphorus studied by angle-resolved photoemission spectroscopy},
	volume = {90},
	url = {https://link.aps.org/doi/10.1103/PhysRevB.90.085101},
	doi = {10.1103/PhysRevB.90.085101},
	pages = {085101},
	number = {8},
	journal = {Physical Review B},
	shortjournal = {Phys. Rev. B},
	author = {Han, C. Q. and Yao, M. Y. and Bai, X. X. and Miao, Lin and Zhu, Fengfeng and Guan, D. D. and Wang, Shun and Gao, C. L. and Liu, Canhua and Qian, Dong and Liu, Y. and Jia, Jin-feng},
	urldate = {2025-03-19},
	date = {2014-08-05},
        year = {2014},
}

@article{faure_full_2012,
	title = {Full characterization and optimization of a femtosecond ultraviolet laser source for time and angle-resolved photoemission on solid surfaces},
	volume = {83},
	issn = {0034-6748},
	url = {https://doi.org/10.1063/1.3700190},
	doi = {10.1063/1.3700190},
	pages = {043109},
	number = {4},
	journal = {Review of Scientific Instruments},
	shortjournal = {Review of Scientific Instruments},
	author = {Faure, J. and Mauchain, J. and Papalazarou, E. and Yan, W. and Pinon, J. and Marsi, M. and Perfetti, L.},
	urldate = {2025-04-08},
	date = {2012-04-13},
        year = {2012},
}

@article{sobota_ultrafast_2012,
	title = {Ultrafast Optical Excitation of a Persistent Surface-State Population in the Topological Insulator {Bi}$_2${Se}$_3$},
	volume = {108},
	url = {https://link.aps.org/doi/10.1103/PhysRevLett.108.117403},
	doi = {10.1103/PhysRevLett.108.117403},
	pages = {117403},
	number = {11},
	journal = {Physical Review Letters},
	shortjournal = {Phys. Rev. Lett.},
	author = {Sobota, J. A. and Yang, S. and Analytis, J. G. and Chen, Y. L. and Fisher, I. R. and Kirchmann, P. S. and Shen, Z.-X.},
	urldate = {2025-04-10},
	date = {2012-03-14},
        year = {2012},
}

@article{rohleder_momentum-resolved_2005,
	title = {Momentum-resolved dynamics of {Ar/Cu(100)} interface states probed by time-resolved two-photon photoemission},
	volume = {7},
	issn = {1367-2630},
	year = {2005},
	url = {https://dx.doi.org/10.1088/1367-2630/7/1/103},
	doi = {10.1088/1367-2630/7/1/103},
	pages = {103},
	number = {1},
	journal = {New Journal of Physics},
	shortjournal = {New J. Phys.},
	author = {Rohleder, M and Duncker, K and Berthold, W and Güdde, J and Höfer, U},
	urldate = {2025-04-29},
	date = {2005-04},
	langid = {english},
}

@article{wolf_femtosecond_1997,
	title = {Femtosecond dynamics of electronic excitations at metal surfaces},
	volume = {377-379},
	issn = {0039-6028},
	year = {1997},
	url = {https://www.sciencedirect.com/science/article/pii/S0039602896014124},
	doi = {10.1016/S0039-6028(96)01412-4},
	series = {European Conference on Surface Science},
	pages = {343--349},
	journal = {Surface Science},
	shortjournal = {Surface Science},
	author = {Wolf, Martin},
	urldate = {2025-04-29},
	date = {1997-04-20},
}

@article{wolf_ultrafast_1996,
	title = {Ultrafast dynamics of electrons in image-potential states on clean and Xe-covered {Cu(111)}},
	volume = {54},
	year = {1996},
	url = {https://link.aps.org/doi/10.1103/PhysRevB.54.R5295},
	doi = {10.1103/PhysRevB.54.R5295},
	pages = {R5295--R5298},
	number = {8},
	journal = {Physical Review B},
	shortjournal = {Phys. Rev. B},
	author = {Wolf, M. and Knoesel, E. and Hertel, T.},
	urldate = {2025-04-29},
	date = {1996-08-15},
}

@article{fauster_femtosecond_2000,
	title = {Femtosecond two-photon photoemission studies of image-potential states},
	volume = {251},
	issn = {0301-0104},
	year = {2000},
	url = {https://www.sciencedirect.com/science/article/pii/S0301010499003006},
	doi = {10.1016/S0301-0104(99)00300-6},
	pages = {111--121},
	number = {1},
	journal = {Chemical Physics},
	shortjournal = {Chemical Physics},
	author = {Fauster, Thomas and Reuß, Christian and Shumay, Igor L and Weinelt, Martin},
	urldate = {2025-04-29},
	date = {2000-01-01},
}

@article{rohleder_time-resolved_2005,
	title = {Time-Resolved Two-Photon Photoemission of Buried Interface States in {Cu(100)}},
	volume = {94},
	year = {2005},
	url = {https://link.aps.org/doi/10.1103/PhysRevLett.94.017401},
	doi = {10.1103/PhysRevLett.94.017401},
	pages = {017401},
	number = {1},
	journal = {Physical Review Letters},
	shortjournal = {Phys. Rev. Lett.},
	author = {Rohleder, M. and Berthold, W. and Güdde, J. and Höfer, U.},
	urldate = {2025-04-29},
	date = {2005-01-11},
}

@article{hertel_ultrafast_1996,
	title = {Ultrafast Electron Dynamics at {Cu(111)}: Response of an Electron Gas to Optical Excitation},
	volume = {76},
	year = {1996},
	url = {https://link.aps.org/doi/10.1103/PhysRevLett.76.535},
	doi = {10.1103/PhysRevLett.76.535},
	shorttitle = {Ultrafast Electron Dynamics at {Cu(111)}},
	pages = {535--538},
	number = {3},
	journal = {Physical Review Letters},
	shortjournal = {Phys. Rev. Lett.},
	author = {Hertel, T. and Knoesel, E. and Wolf, M. and Ertl, G.},
	urldate = {2025-04-29},
	date = {1996-01-15},
}

@phdthesis{link_femtosecond_2001,
	title = {Femtosecond electron dynamics of image-potential states on the transition-metal surfaces of Pt and Ni},
	url = {https://www.osti.gov/etdeweb/biblio/20230951},
	author = {Link, Sven},
	urldate = {2025-04-29},
	date = {2001-06-01},
    school = {ForschungsZentrum Jülich},
    year = 2001
}

@article{hofer_time-resolved_1997,
	title = {Time-Resolved Coherent Photoelectron Spectroscopy of Quantized Electronic States on Metal Surfaces},
	volume = {277},
        year = {1997},
	url = {https://www.science.org/doi/full/10.1126/science.277.5331.1480},
	doi = {10.1126/science.277.5331.1480},
	pages = {1480--1482},
	number = {5331},
	journal = {Science},
	author = {Höfer, U. and Shumay, I. L. and Reuß, Ch. and Thomann, U. and Wallauer, W. and Fauster, Th.},
	urldate = {2025-04-30},
	date = {1997-09-05},
}

@article{berthold_momentum-resolved_2002,
	title = {Momentum-Resolved Lifetimes of Image-Potential States on {Cu(100)}},
	volume = {88},
        year = {2002},
	url = {https://link.aps.org/doi/10.1103/PhysRevLett.88.056805},
	doi = {10.1103/PhysRevLett.88.056805},
	pages = {056805},
	number = {5},
	journal = {Physical Review Letters},
	shortjournal = {Phys. Rev. Lett.},
	author = {Berthold, W. and Höfer, U. and Feulner, P. and Chulkov, E. V. and Silkin, V. M. and Echenique, P. M.},
	urldate = {2025-05-01},
	date = {2002-01-18},
}

@article{echenique_decay_2004,
	title = {Decay of electronic excitations at metal surfaces},
	volume = {52},
        year = {2004},
	doi = {10.1016/j.surfrep.2004.02.002},
	pages = {219--317},
	journal = {Surface Science Reports - {SURF} {SCI} {REP}},
	shortjournal = {Surface Science Reports - {SURF} {SCI} {REP}},
	author = {Echenique, P. and Berndt, R. and Chulkov, E. and Fauster, Thomas and Goldmann, A. and Höfer, Ulrich},
	date = {2004-05-01},
}

@article{PhysRevB.111.235109,
  title = {Conduction band structure and ultrafast dynamics of ferroelectric $\ensuremath{\alpha}\text{\ensuremath{-}}\mathrm{GeTe}(111)$},
  author = {Kremer, Geoffroy and Nicola\"{\i}, Laurent and Chassot, Fr\'ed\'eric and Maklar, Julian and Nicholson, Christopher W. and Dil, J. Hugo and Krempask\'y, Juraj and Springholz, Gunther and Ernstorfer, Ralph and Min\'ar, Jan and Rettig, Laurenz and Monney, Claude},
  journal = {Phys. Rev. B},
  volume = {111},
  issue = {23},
  pages = {235109},
  numpages = {9},
  year = {2025},
  month = {Jun},
  publisher = {American Physical Society},
  doi = {10.1103/PhysRevB.111.235109},
  url = {https://link.aps.org/doi/10.1103/PhysRevB.111.235109}
}

@article{elmers_spin_2016,
	title = {Spin mapping of surface and bulk Rashba states in ferroelectric $\alpha$ -{GeTe}(111) films},
	volume = {94},
	rights = {http://link.aps.org/licenses/aps-default-license},
	issn = {2469-9950, 2469-9969},
	url = {https://link.aps.org/doi/10.1103/PhysRevB.94.201403},
	doi = {10.1103/PhysRevB.94.201403},
	pages = {201403},
	number = {20},
	journal = {Physical Review B},
	shortjournal = {Phys. Rev. B},
	author = {Elmers, H. J. and Wallauer, R. and Liebmann, M. and Kellner, J. and Morgenstern, M. and Wang, R. N. and Boschker, J. E. and Calarco, R. and Sánchez-Barriga, J. and Rader, O. and Kutnyakhov, D. and Chernov, S. V. and Medjanik, K. and Tusche, C. and Ellguth, M. and Volfova, H. and Borek, St. and Braun, J. and Minár, J. and Ebert, H. and Schönhense, G.},
	urldate = {2025-08-28},
	date = {2016-11-09},
        year = {2016},
	langid = {english},
}

@article{varotto_room-temperature_2021,
	title = {Room-temperature ferroelectric switching of spin-to-charge conversion in germanium telluride},
	volume = {4},
	rights = {2021 The Author(s), under exclusive licence to Springer Nature Limited},
	issn = {2520-1131},
	url = {https://www.nature.com/articles/s41928-021-00653-2},
	doi = {10.1038/s41928-021-00653-2},
	pages = {740--747},
	number = {10},
	journal = {Nature Electronics},
	shortjournal = {Nat Electron},
	author = {Varotto, Sara and Nessi, Luca and Cecchi, Stefano and Sławińska, Jagoda and Noël, Paul and Petrò, Simone and Fagiani, Federico and Novati, Alessandro and Cantoni, Matteo and Petti, Daniela and Albisetti, Edoardo and Costa, Marcio and Calarco, Raffaella and Buongiorno Nardelli, Marco and Bibes, Manuel and Picozzi, Silvia and Attané, Jean-Philippe and Vila, Laurent and Bertacco, Riccardo and Rinaldi, Christian},
	urldate = {2025-08-28},
	date = {2021-10},
        year = {2021},
	langid = {english},
}

@article{krempasky2020fully,
  title={Fully spin-polarized bulk states in ferroelectric GeTe},
  author={Krempask{\`y}, Juraj and Fanciulli, Mauro and Nicola{\"\i}, Laurent and Min{\'a}r, Jan and Volfov{\'a}, Henrieta and Caha, Ond{\v{r}}ej and Volobuev, Valentine V and S{\'a}nchez-Barriga, Jaime and Gmitra, Martin and Yaji, Koichiro and others},
  journal={Physical Review Research},
  volume={2},
  number={1},
  pages={013107},
  year={2020},
  publisher={APS}
}

@article{echenique1978existence,
  title={The existence and detection of Rydberg states at surfaces},
  author={Echenique, PM and Pendry, JB},
  journal={Journal of Physics C: Solid State Physics},
  volume={11},
  number={10},
  pages={2065},
  year={1978},
  publisher={IOP Publishing}
}

@article{straub1984intrinsic,
  title={Intrinsic unoccupied surface states at GaP (110)},
  author={Straub, D and Altmann, W and Scheidt, H and Dose, V},
  journal={Journal of Vacuum Science \& Technology A: Vacuum, Surfaces, and Films},
  volume={2},
  number={2},
  pages={529--530},
  year={1984},
  publisher={American Vacuum Society}
}

@article{dose1984image,
  title={Image-potential states observed by inverse photoemission},
  author={Dose, V and Altmann, W and Goldmann, A and Kolac, U and Rogozik, J},
  journal={Physical review letters},
  volume={52},
  number={21},
  pages={1919},
  year={1984},
  publisher={APS}
}

@article{tognolini2015rashba,
  title={Rashba spin-orbit coupling in image potential states},
  author={Tognolini, Silvia and Achilli, Simona and Longetti, L and Fava, E and Mariani, Carlo and Trioni, MI and Pagliara, Stefania},
  journal={Physical review letters},
  volume={115},
  number={4},
  pages={046801},
  year={2015},
  publisher={APS}
}

@article{nakazawa2016rashba,
  title={Rashba splitting in an image potential state investigated by circular dichroism two-photon photoemission spectroscopy},
  author={Nakazawa, T and Takagi, N and Kawai, Maki and Ishida, H and Arafune, R},
  journal={Physical Review B},
  volume={94},
  number={11},
  pages={115412},
  year={2016},
  publisher={APS}
}

@article{schottke2022rashba,
  title={Rashba-split image-potential state and unoccupied surface electronic structure of Re (0001)},
  author={Sch{\"o}ttke, Fabian and Schemmelmann, Sven and Kr{\"u}ger, Peter and Donath, Markus},
  journal={Physical Review B},
  volume={105},
  number={15},
  pages={155419},
  year={2022},
  publisher={APS}
}

@article{braun2021impact,
  title={The Impact of Spin--Orbit Interaction on the Image States of High-Z Materials},
  author={Braun, J{\"u}rgen and Ebert, Hubert},
  journal={physica status solidi (b)},
  volume={258},
  number={1},
  pages={2000026},
  year={2021},
  publisher={Wiley Online Library}
}

@article{Vosko_1980,
author = {Vosko, S. H. and Wilk, L. and Nusair, M.},
title = {Accurate spin-dependent electron liquid correlation energies for local spin density calculations: a critical analysis},
journal = {Canadian Journal of Physics},
volume = {58},
number = {8},
pages = {1200-1211},
year = {1980},
doi = {10.1139/p80-159},
URL = {https://doi.org/10.1139/p80-159},
}

@article{moser2018extract,
  title={How to extract the surface potential profile from the ARPES signature of a 2DEG},
  author={Moser, S and Jovic, V and Koch, R and Moreschini, L and Oh, J-S and Jozwiak, C and Bostwick, A and Rotenberg, E},
  journal={Journal of Electron Spectroscopy and Related Phenomena},
  volume={225},
  pages={16--22},
  year={2018},
  publisher={Elsevier}
}

@article{tsu1968optical,
  title={Optical and electrical properties and band structure of GeTe and SnTe},
  author={Tsu, R and Howard, W Ef and Esaki, L},
  journal={Physical Review},
  volume={172},
  number={3},
  pages={779},
  year={1968},
  publisher={APS}
}

@article{chen2017dielectric,
  title={Dielectric properties of amorphous phase-change materials},
  author={Chen, Chao and Jost, Peter and Volker, Hanno and Kaminski, Marvin and Wirtssohn, M and Engelmann, U and Kr{\"u}ger, K and Schlich, F and Schlockermann, Carl and Lobo, Ricardo PSM and others},
  journal={Physical Review B},
  volume={95},
  number={9},
  pages={094111},
  year={2017},
  publisher={APS}
}

@article{miaja2006laser,
  title={Laser-assisted photoelectric effect from surfaces},
  author={Miaja-Avila, Luis and Lei, C and Aeschlimann, M and Gland, JL and Murnane, MM and Kapteyn, HC and Saathoff, G},
  journal={Physical review letters},
  volume={97},
  number={11},
  pages={113604},
  year={2006},
  publisher={APS}
}

@article{giesen1987effective,
  title={Effective mass of image-potential states},
  author={Giesen, K and Hage, F and Himpsel, FJ and Riess, HJ and Steinmann, W and Smith, NV},
  journal={Physical Review B},
  volume={35},
  number={3},
  pages={975},
  year={1987},
  publisher={APS}
}

@article{ponzoni2023dirac,
  title={Dirac bands in the topological insulator Bi2Se3 mapped by time-resolved momentum microscopy},
  author={Ponzoni, Stefano and Pa{\ss}lack, Felix and Stupar, Matija and Janas, David Maximilian and Zamborlini, Giovanni and Cinchetti, Mirko},
  journal={Advanced Physics Research},
  volume={2},
  number={5},
  pages={2200016},
  year={2023},
  publisher={Wiley Online Library}
}

@article{ferrini2003effective,
  title={Effective mass and momentum-resolved intrinsic linewidth of image-potential states on Ag (100)},
  author={Ferrini, Gabriele and Giannetti, Claudio and Fausti, Daniele and Galimberti, Gianluca and Peloi, Marco and Banfi, GianPiero and Parmigiani, Fulvio},
  journal={Physical Review B},
  volume={67},
  number={23},
  pages={235407},
  year={2003},
  publisher={APS}
}

@article{echenique2002image,
  title={Image-potential-induced states at metal surfaces},
  author={Echenique, PM and Pitarke, JM and Chulkov, EV and Silkin, VM},
  journal={Journal of electron spectroscopy and related phenomena},
  volume={126},
  number={1-3},
  pages={163--175},
  year={2002},
  publisher={Elsevier}
}
\end{document}